\documentclass[twocolumn,showpacs,preprintnumbers,amsmath,amssymb]{revtex4}

\usepackage{graphicx}
\usepackage{dcolumn}
\usepackage{bm}

\begin{document}

\title{An Atom Laser is not monochromatic}
\author{S. Choi$^{1}$,  D. Str\"{o}mberg$^{1,2}$,
and B. Sundaram$^{1}$} \affiliation{$^{1.}$ Department of Physics,
University of Massachusetts, Boston, MA 02125, USA \\  $^{2.}$
Department of Information Technology, Uppsala University, 751 05
Uppsala, Sweden}

\begin{abstract}
We study both numerically and analytically the possibility of using
an adiabatic passage control method to construct a Mach-Zehnder
interferometer (MZI) for Bose-Einstein condensates (BECs) in the
time domain, in exact one-to-one correspondence with the
traditional optical MZI that involves two beam splitters and two
mirrors. The interference fringes one obtains from such a
minimum-disturbance set up clearly demonstrates that, fundamentally,
an atom laser is not monochromatic due to interatomic interactions.
We also consider how the amount of entanglement in the system
correlates to the interference fringes.
\end{abstract}


\maketitle

Atomic Bose-Einstein condensates (BECs) are often referred to as the
``atom laser,'' the matter-wave equivalent of a laser. Unlike
massless photons from a laser  which do not interact, atoms from an
atom laser do interact with each other, and so far it has not been
clearly established how and to what extent the monochromaticity of
an atom laser is compromised by such interactions. The term ``atom
laser'' is not yet defined universally other than in a broad general
sense, although there have been attempts to be more
specific about what an atom laser should be, including the
directionality of the beam in a close analogy with a
laser~\cite{Wiseman}. However, given fundamental
differences such as the fact that
the atoms in a BEC are intrinsically stationary, here we shall use
the term to denote an {\it intense} source of {\it
coherent} atoms  {\it for atom optical transformations} such as
reflection, refraction and beam splitting.

In general, the non-zero
mass of the atoms necessitates rather elaborate and experimentally
demanding schemes to coherently accelerate and guide the
matter-wave. In order to mitigate this problem, we propose the use
of atom optics in the {\it time-domain} based on adiabatic passage.
An important reason for this is that time domain atom optics is less
prone to decoherence due to, say, atom losses as compared to
techniques such as guiding a BEC in space. Further, the
adiabatic passage method is a well-known technique in quantum
mechanics that provides a way to ``delicately'' control a quantum
system via slow passage along the energy landscape, while preserving
the quantum system in one of the energy eigenstates, typically the
ground state.  By minimizing any disturbance, and maintaining the
condensate in the ground state, one can draw conclusions about the
intrinsic properties of an atom laser unmodified by such additional
actions as ballistic expansion.

\begin{figure}
\begin{center}
\centerline{\includegraphics[height=10.5cm]{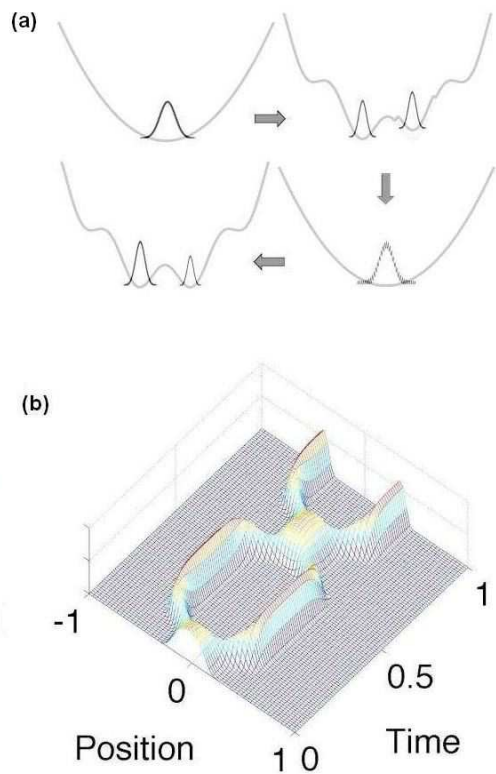}}
 \caption{(a) Schematic diagram of the
transformations that the potential undergoes over time, starting
with beam splitting, addition of a potential step, recombination and
final beam splitting.  (b) Spatio-temporal probability density obtained by solving the
Gross-Pitaevkii equation describing the condensate passing through
the time-domain MZI.
 } \label{fig1}
\end{center}
\vspace{-0.5in}
\end{figure}

To address the question of the monochromaticity of an atom laser, we
model both numerically and analytically an ideal Mach-Zehnder
Interferometer (MZI) for BECs in the time domain and deduce the
spectral composition of an atom laser from the interference pattern
generated. The construction of a MZI that uses BEC as the source of
coherent atoms is one of the goals in atom optics owing to
the promise of, for instance, high precision matter-wave based
metrology. In a typical optical MZI scheme, coherent light from a
laser is passed through a beam splitter which sends the light beam
into two paths; one of the beams passes through a phase shifter and
the two beams are then recombined by reflecting off mirrors onto a
second beam splitter. The phase shift experienced in one arm is
reflected in the intensity variation at the two detectors. We have
simulated the time-domain MZI by adding an optical potential to the
usual quadratic trapping potential and adiabatically varying its
amplitude to simulate exactly the action of the optical elements in
an usual MZI, namely  the two beam splitters, mirrors and wave
guides. The phase shift is provided by adding an additional
potential step to one ``arm'' of the interferometer. Given the
adiabatic nature of the method, the two arms do not interfere unless
they are explicitly recombined by way of the second beam splitter.
The population difference $\Delta P$ between the two arms is finally
measured as a function of the step height.  A schematic diagram of
the required operations is provided in Fig. \ref{fig1}(a). To the
best of our knowledge, such a scheme has not yet been experimentally
implemented, although there have been several experiments on
interference of BEC in optical potentials~\cite{DoubleWell} and in
atom chips~\cite{AtomChip,AtomChip2, AtomChip3}, with the important
difference that all of these release the condensates to expand
ballistically after the first beam splitter. The time-domain MZI
proposed here keeps the condensate in the ground state, better
revealing the nature of the atom laser.

\begin{figure}
\begin{center}
\centerline{\includegraphics[height=6cm]{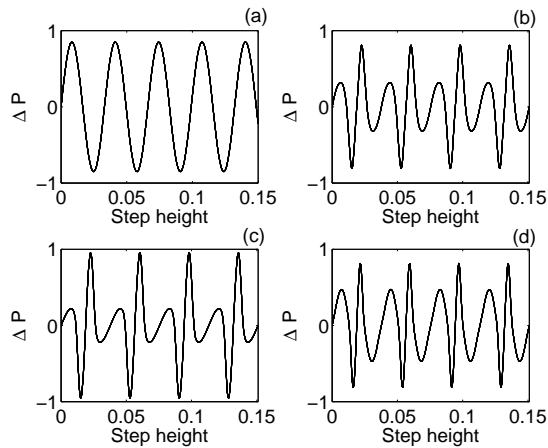}}
\caption{Interference fringes obtained by solving the
Gross-Pitaevskii Equation (GPE) to model the time-domain MZI for
various values of the nonlinearity constant $g$ {\bf (a)} $g = 0$
{\bf (b)} $g = 0.5$ {\bf (c)} $g = 5$ {\bf (d)} $g = 10$ The
sinusoidal fringe for the linear $g=0$ case is clearly modified for
$g
> 0$ i.e. in the presence of interatomic interactions, showing that the atom laser is
no longer monochromatic.} \label{fig2}
\end{center}
\vspace{-0.4in}
\end{figure}

The Gross-Pitaevskii Equation (GPE) was used to simulate the
dynamics of the condensate in the presence of the spatio-temporal
potential. The GPE has been extremely
successful in describing the dynamics of BECs under a variety of
experimental conditions including optical lattices\cite{Review}.
Specifically, we consider the 1-D GPE:
\begin{equation}
i\hbar \frac{\partial \psi}{\partial t} = \left [
-\frac{\hbar^2}{2m} \frac{d^2}{dx^2} + V(x,t) + g|\psi(x,t)|^2
\right ] \psi(x,t)
\end{equation}
where $\psi(x,t)$ denotes the condensate mean field and $V(x,t)$
denotes the adiabatically changing time-dependent potential. As
shown later in our analysis, the dimensionality of
the GPE simulation does not affect the final results. We always
retain the parabolic trapping potential, so $V(x,t) =
\frac{1}{2} m \omega_{t} x^2 + Af(t) \cos (k x) + V_{step}(x,
T_{1/4}<t<2T_{1/4})$ where $k = 2\pi/ \lambda$, $A$ is the amplitude
of the optical lattice potential
 and $T_{1/4}$ denotes $1/4$ of the
total duration of the simulation $T$. $f(t)$ provides the time
dependence for the adiabatically changing amplitude of the optical
lattice:
\begin{equation}
f(t) = \left\{ \begin{array}{c l}
         t/T_{1/4} &  \mbox{  $0 < t < T_{1/4}$, first beam splitter} ;\\
          1 &  \mbox{  $T_{1/4} < t < 2T_{1/4}$, first beam splitter} ; \\
         (3 - t/T_{1/4}) &  \mbox{  $2T_{1/4} < t < 3T_{1/4}$, recombination}; \\
         (t/T_{1/4} - 3) &  \mbox{  $3T_{1/4} < t < 4T_{1/4}$, final beam splitter}.\end{array}
\right.
\end{equation}
The potential step was introduced during $T_{1/4} < t < 2T_{1/4}$
using a super-Gaussian  in one arm $V_{step}(x, t) =  h e^{ - [(t -
T_{1/4})/\sigma_{t}]^{10}} e^{ - [(x -
\lambda/2)/\sigma_{x}]^{30}}$, with the  spatial and temporal width
of $\sigma_{x}  = \lambda/3$ and $\sigma_{t}  = T_{1/4}/2$
respectively.  The probability density of the condensate obtained
from GPE dynamics with $g = 5$, $A = 25$, $T_{1/4} = 400$, and $\lambda = 15$
in harmonic oscillator units of the initial trapping potential is
shown in Fig. \ref{fig1}(b) as a spatio-temporal plot. As is clear
from the figure, the simulation provides exact one-to-one
correspondence to the MZI, and the adiabatic condition ensures that the
wave function evolves extremely cleanly.  One of the advantages of
this scheme of interferometry is that the area enclosed by the two
arms of the interferometer is readily controllable by changing
$\lambda$; in particular, with a larger enclosed area, higher
sensitivity is possible for use in, for instance, interferometric
rotational sensors used in space navigation.

We plot in Fig. \ref{fig2} the
final intensity difference as a function of step height $h$ for
different values of nonlinearity $g$. It is noted that a relatively
large phase shift  is recorded even for small variations in step
height, indicating high sensitivity of the condensate to changes in
potential height. In the linear case (non BEC) $g = 0$ one obtains
perfect sinusoidal variation as a function of $h$, while for $g >
0$, the pattern clearly deviates from the sinusoidal, implying that
an atom laser loses monochromaticity due to the presence of
interatomic interactions. This is confirmed by a Fourier transform
which reveals four closely spaced frequency components.

In order to better understand this result and to ensure that the
adiabatic assumptions are not violated, a detailed analysis of this
system can be given as follows. First, we assume that the adiabatic
passage we used above preserves the single mode approximation. The
condensate after beam splitting and the potential step (which
imparts relative phase of $\Theta$) may then be written as:
\begin{equation}
\hat{\psi}(x, t, \Theta) = \phi_{L}(x,d) \hat{a}_{L}(t) +
\phi_{R}(x,d) \exp(i \Theta) \hat{a}_{R} (t)  \label{singlemode}
\end{equation}
where $\hat{a}_{L(R)}(t)$ are bosonic annihilation operators for the
modes of the matter wave field in the left and right arms and
$\phi_{L(R)}(x,d)$ are the corresponding spatial mode functions.
These may be approximated by spatial functions such as shifted
Gaussians $\phi_{L(R)}(x,d) = (\pi \sigma^2)^{-1/4} e^{-(x \pm
d)^2/4 \sigma^2}$ that, under adiabatic evolution, do not change
shape. The phase shift has been modeled by the explicit insertion of
the $\exp(i \Theta)$ term, which is clearly an approximation to the
full GPE simulation where the phase shift results from the passage
of the condensate over a potential step in one arm. The analysis is
therefore best suited to the cases with lower values of $g$ for
which interplay between the potential step and repulsive atomic
interactions do not complicate the phase relationship given in Eq.
(\ref{singlemode}).

Since we use the adiabatic passage method, there are no additional
net physical effects on the BEC undergoing the remaining two operations that
constitute the MZI (recombination and then second beam splitting),
other than those of the interatomic scattering
and the natural tunnelling flow between the left and right arms. This
is due to the Josephson effect and the geometry of the system though
the tunnelling would be minimized when the arms are well separated.
The effective interference fringes are then given by the
difference in the number of atoms in the two arms after the system
has evolved a certain time $\tau$ through the MZI.  With the field
decomposition of Eq. (\ref{singlemode}) the bosonic Hamiltonian is given
by
\begin{eqnarray}
\hat{H} & = & \hbar \omega (\hat{a}^{\dagger}_{L}\hat{a}_{L} +
\hat{a}^{\dagger}_{R}\hat{a}_{R}) + \frac{\Delta E(t)}{2}
[\hat{a}^{\dagger}_{L}\hat{a}_{R} +
\hat{a}^{\dagger}_{R}\hat{a}_{L}] \nonumber \\
& & + g(\hat{a}^{\dagger 2}_{L}\hat{a}^{2}_{L} + \hat{a}^{\dagger
2}_{R}\hat{a}^{2}_{R})
\end{eqnarray}
where $\Delta E(t) = \hbar \omega \exp [-d^{2}(t)/\sigma^2]$ gives
the overlap of the two spatial modes and is the time-dependent
tunneling energy.  This overlap clearly depends on the
dimensionality of the modes. However, change in dimensionality
merely alters this single scaling parameter and not the behavior we
describe. This Hamiltonian is also well-known as the Hamiltonian
that describes two-component BECs when ``L'' and ``R'' denote two
different species instead of the left and right arms. Introducing
the Schwinger angular
momentum operators, $\hat{J}_{+(-)} = \hat{a}_{L(R)}^{\dagger}\hat{a}_{R(L)}$ and $\hat{J}%
_z = \frac{1}{2} ( \hat{a}_{L}^{\dagger}\hat{a}_{L} -
\hat{a}_{R}^{\dagger}\hat{a}_{R})$, the Hamiltonian is equivalent to
a ``Lipkin-Meshkov-Glick'' type Hamiltonian~\cite{LMG},
\begin{eqnarray}
\hat{H} & = &  \frac{\Delta E(t)}{2} \left ( \hat{J}_{+} e^{i
\Theta} + \hat{J}_{-} e^{-i \Theta} \right ) + 2g \hat{J}_{z}^2 .
\end{eqnarray}
The time evolution operator corresponding
to this nonlinear Hamiltonian has been studied
previously~\cite{ChoiEntangle}.   The measured intensity difference
is given by $\langle J_{z} (\tau) \rangle$ at time $\tau$,
$\langle \hat{J}_{z} (\tau, \Theta) \rangle   =  \langle
\hat{R}^{\dagger}e^{iH^{\prime} \tau }\hat{R}
 \hat{J}_{z}  \hat{R}^{\dagger}e^{-iH^{\prime} \tau }\hat{R} \rangle$,
 where $\hat{R} =
\exp \left [ -\frac{\pi}{4} (\hat{J}_{-} e^{-i \Theta} - \hat{J}_{+}
e^{i \Theta}) \right ]$ is the well-known rotation operator and $H'$
is the transformed Hamiltonian diagonal in $\hat{J}_{z}$:
$\hat{H}^{\prime}  =\hat{R}\hat{H} \hat{R}^{\dagger} - i \hat{R} \frac{%
\partial}{\partial t} \hat{R}^{\dagger}  \equiv \Delta E \hat{J}_{z} - g
\hat{J}_{z}^{2}$.

Writing the quantum state in a general form $| \Psi \rangle =
\sum_{m} c_m |m \rangle_{L} | N- m \rangle_{R}$ where $N$ is the
total number of atoms in the system, one obtains,
\begin{eqnarray}
\langle \hat{J}_{z} (\tau, \Theta) \rangle
 & = & - \frac{e^{i g\tau }}{2}  \left ( \sum_{m'}  \beta^{(+)}_{m'} \gamma^{(+)*}_{m'}  \right ) e^{i (\Theta + \Delta \omega \tau)} \nonumber \\
& &  - \frac{e^{i g\tau }}{2}  \left (\sum_{m'}  \beta^{(-)}_{m'}
\gamma^{(-)*}_{m'} \right ) e^{-i(\Theta + \Delta \omega  \tau)} \nonumber
\end{eqnarray}
where $\Delta \omega  = \Delta E /\hbar$ and
\begin{eqnarray}
\beta^{(\pm)}_{m'}
& \equiv  & \sum_{m} c^{*}_{m} {\cal R}_{m,m'} \sqrt{j(j + 1) - m'(m' \pm 1)}  \nonumber \\
\gamma^{(\pm)}_{m'}
& \equiv &   \sum_{m} c^{*}_{m} {\cal R}_{m,m' \pm 1} e^{\pm i 2 m'
g \tau }
\end{eqnarray}
with the matrix elements ${\cal R}_{m,m'} \equiv \langle m |
\hat{R}^{\dagger} | m' \rangle$ provided by an analytical
expression~\cite{bloch}.

\begin{figure}
\begin{center}
\centerline{\includegraphics[height=5cm]{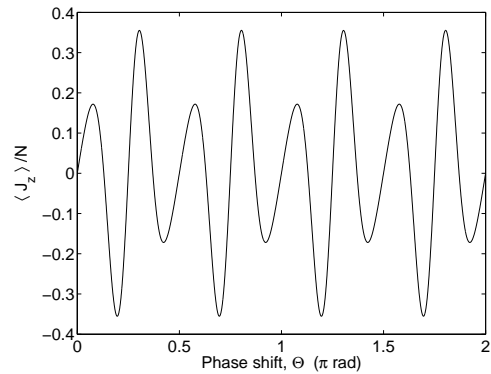}} \caption{ The
interference fringe given by the fractional number difference in the
two arms of the interferometer, $\langle \hat{J}_{z} \rangle/N$,
obtained using the analytical methods under the adiabatic
assumption. The figure, which shows remarkable agreement with the
simulation results obtained above using the GPE, again clearly
indicates the presence of several frequency components in the atom
laser from the way it deviates from the sinusoidal pattern.}
\label{fig3}
\end{center}
\vspace{-0.4in}
\end{figure}

A natural choice of quantum state for the two component BECs with a
relative phase $\Theta$  is a coherent spin state or the Bloch state
with equal number of atoms in both arms such that $|\Psi \rangle =
|\theta = \pi/2, \phi = \Theta \rangle$. In this case, the
coefficients $c_m = (C^{2j}_{j + m})^{1/2} e^{i(j-m)\Theta}/2^j$
where $C^{n}_{m}$ denotes the combination, $C^{n}_{m} =
n!/[(n-m)!m!]$. The resulting interference fringe with $(\Delta
\omega) \tau$ chosen to be $(2n + 2)\pi$  and $g\tau = (2n +
\frac{3}{2N})\pi$, where $n$ is an integer, is plotted in Fig.
\ref{fig3} which shows remarkable agreement with the result from the
GPE simulation. With $n \approx 127$, we are simulating the $g=0.5$
case in the GPE simulation above, and the effective magnitude of $g$
of the order $\frac{3}{2N}$ places the system in the Josephson
regime. The amplitudes are different from the GPE simulation simply
due to our choice of $\langle \hat{J}_{z} \rangle$ to be half the
atom number difference.

The results confirm that the coherent spin state is a natural state
for BECs with relative phase arising from the adiabatic passage
method. With the current system, the initial state prepared at the
first beam splitter can be controlled so that the number of atoms in
each arm are not equally distributed.  For instance, this can be accomplished by
adiabatically deforming the potential wells asymmetrically to collect
more atoms in one arm and then continue adiabatically to model
subsequent actions such as recombination. In such cases, a coherent spin state with
$\theta \neq \pi/2$ can be generated, and consequently the amount of
entanglement between the left and right arms is directly
controllable. The state with $\theta = \pi/2$ has the largest entanglement
while the state with $\theta = 0$ has the least amount of entanglement~\cite{ChoiEntangle}.

\begin{figure}
\begin{center}
\centerline{\includegraphics[height=6cm]{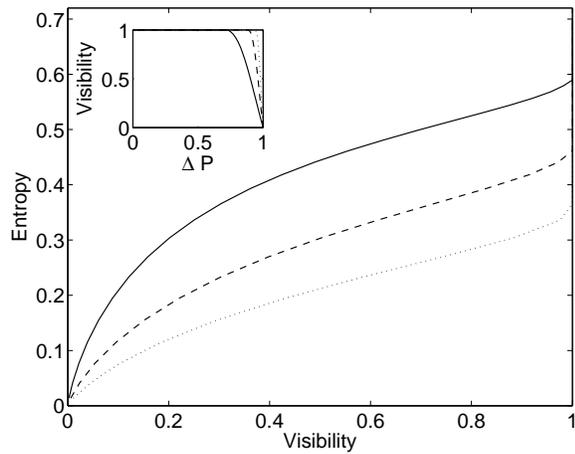}}
\caption{Entanglement parameterized by the von Neumann entropy of
the input state plotted against the visibility of the final
inference fringes. Inset: Visibility vs. fractional atom number
difference between the two arms $\Delta P$ of the input state.
Solid, dashed, and dotted lines: equivalent nonlinearity in GPE of
$g = 0$, $0.25$ and $0.5$ respectively in both the main figure and
the inset. } \label{entangle}
\end{center}
\vspace{-0.4in}
\end{figure}

Finally, how the presence of interatomic interaction modifies the
connection between the fringe visibility  and entanglement is of
interest from the point of possible experimental determination of
parameters such as the interatomic scattering length using this
method. The degree of entanglement between the left and right arms
can be quantified using the von Neumann entropy
$E(t) = -\frac{1}{\log_2(N+1)} \sum_{m = -j}^{j} |c_m|^2 \log_2
|c_m|^2\;,$
which is clearly connected to the interference pattern at time
$\tau$ given by $\langle \hat{J}_z (\tau, \Theta) \rangle$. We plot
in Fig. \ref{entangle} the von Neumann entropy of the input state
against visibility ${\cal V}$ of the fringes ${\cal V} = (I_{max} -
I_{min})/(I_{max} + I_{min})$ where $I= |\langle \hat{J}_{z}(\tau)
\rangle|^2$ denotes the intensity of the fringe pattern. The
visibility of the interference fringes is obviously reduced with an
asymmetric beam splitting that corresponds to $\theta \neq \pi/2$.
The different line shapes correspond to different values of
nonlinearity. For a given value of visibility  one obtains lower
entanglement with higher nonlinearity. This initially
counterintuitive result can be understood from the inset of Fig.
\ref{entangle}, which provides the relationship between the
fractional number difference between the two arms of the initial
input state $\Delta P$ and the visibility of the fringes. As
expected ${\cal V} \rightarrow 0$ as $\Delta P \rightarrow 1$. For
$\Delta P < 0.75$, ${\cal V}$ already reaches 1 as one starts
finding atoms in the other arm such that $\langle \hat{J}_{z}
\rangle$ starts to take on negative values as a function of
$\Theta$. With higher nonlinearity one finds ${\cal V} \rightarrow
1$ even for larger values of $\Delta P$. This may be understood as
the result of higher entanglement due to collisions. With higher
entanglement, the ``wave-like'' nature of the wave function is
enhanced (changing one arm affects the other arm instantaneously)
which encourages tunneling between the two arms, and subsequently
higher visibility.

In conclusion, we have studied a potentially realizable scheme for
a MZI using BECs. The resulting intensity difference between the
two arms as a function of phase shift is found to demonstrate that,
fundamentally, an atom laser is not monochromatic but rather
comprises of a number of frequency components.  The adiabatic
passage method ensured that the only ``alteration'' made to the
condensate is that of coherent splitting and the establishment of a
relative phase between the two arms which naturally puts the trapped
condensate onto a coherent spin state. The dynamics of the split
condensate was then given simply by a standard many-body Hamiltonian
over a finite time interval. The fact that monochromaticity is lost
even in this idealized model implies that one should exercise
caution when discussing an atom laser as simply a matter-wave
equivalent of a laser, since additional effects such as decoherence
due to atom losses is expected to degrade the quality even further.
Finally we observed that with higher nonlinearity, higher fringe
visibility can be obtained with very uneven beam splitting owing to
the enhanced entanglement between the two arms. This work suggests
that there are many new aspects to be explored in the time domain
quantum control of BEC using adiabatic passage.

\end{document}